\documentclass[twocolumn,english,aps,prb,superscriptaddress]{revtex4-1}
\usepackage{babel}
\usepackage{amsmath}
\usepackage{graphicx}
\usepackage{amssymb}
\usepackage{color}
\usepackage[dvipsnames]{xcolor}
\usepackage{hyperref}

\hypersetup{
    colorlinks,
    linkcolor={blue},
    citecolor={blue},
    urlcolor={blue}
}

\newcommand{\aio}{$A_2$IrO$_3$}
\newcommand{\rucl}{$\alpha$-RuCl$_3$}

\newcommand{\cac}{\chi'_{\rm ac}}

\newcommand{\lllangle}{\langle\!\langle\!\langle}
\newcommand{\rrrangle}{\rangle\!\rangle\!\rangle}

\newcommand{\TN}{T_{\rm N}}

\graphicspath{{./}{./plots/}}

\begin{document}

\title{
Field-induced intermediate phase in \rucl: \\
Non-coplanar order, phase diagram, and proximate spin liquid
}

\author{P. Lampen-Kelley\cite{contrib, plk}}
\affiliation{Department of Materials Science and Engineering, University of Tennessee, Knoxville, TN 37996, U.S.A.}
\affiliation{Materials Science and Technology Division, Oak Ridge National Laboratory, Oak Ridge, TN 37831, U.S.A.}

\author{L. Janssen\cite{contrib, lj}}
\affiliation{Institut f\"ur Theoretische Physik, Technische Universit\"at Dresden,
01062 Dresden, Germany}

\author{E. C. Andrade}
\affiliation{Instituto de F\'{i}sica de S\~ao Carlos, Universidade de S\~ao Paulo, C.P. 369,
S\~ao Carlos, SP 13560-970, Brazil}

\author{S. Rachel}
\affiliation{School of Physics, University of Melbourne, Parkville, VIC 3010, Australia}

\author{J.-Q.~Yan}
\affiliation{Materials Science and Technology Division, Oak Ridge National Laboratory, Oak Ridge, TN 37831, U.S.A.}

\author{C. Balz}
\affiliation{Neutron Scattering Division, Oak Ridge National Laboratory, Oak Ridge, TN 37831, U.S.A.}

\author{D. G. Mandrus}
\affiliation{Department of Materials Science and Engineering, University of Tennessee, Knoxville, TN 37996, U.S.A.}
\affiliation{Materials Science and Technology Division, Oak Ridge National Laboratory, Oak Ridge, TN 37831, U.S.A.}

\author{S. E. Nagler}
\affiliation{Neutron Scattering Division, Oak Ridge National Laboratory, Oak Ridge, TN 37831, U.S.A.}

\author{M. Vojta}
\affiliation{Institut f\"ur Theoretische Physik, Technische Universit\"at Dresden,
01062 Dresden, Germany}


\maketitle


\textbf{
Frustrated magnets with strong spin-orbit coupling promise to host topological states of matter, with fractionalized excitations and emergent gauge fields. Kitaev's proposal for a honeycomb-lattice Majorana spin liquid \cite{Kit06} has triggered an intense search for experimental realizations, with bond-dependent Ising interaction being the essential building block. A prime candidate is $\alpha$-RuCl$_3$ whose phase diagram in a magnetic field is, however, not understood to date.
Here we present conclusive experimental evidence for a novel field-induced ordered phase in $\alpha$-RuCl$_3$, sandwiched between the zigzag and quantum disordered phases at low and high fields, respectively. We provide a detailed theoretical study of the relevant effective spin model which we show to display a field-induced intermediate phase as well. We fully characterize the intermediate phase within this model, including its complex spin structure, and pinpoint the parameters relevant to $\alpha$-RuCl$_3$ based on the experimentally observed critical fields.
Most importantly, our study connects the physics of $\alpha$-RuCl$_3$ to that of the Kitaev-$\Gamma$ model, which displays a quantum spin liquid phase in zero field, and hence reveals the spin liquid whose signatures have been detected in a variety of dynamical probes of $\alpha$-RuCl$_3$.
}

\begin{figure}[b]
\includegraphics[width=0.38\textwidth]{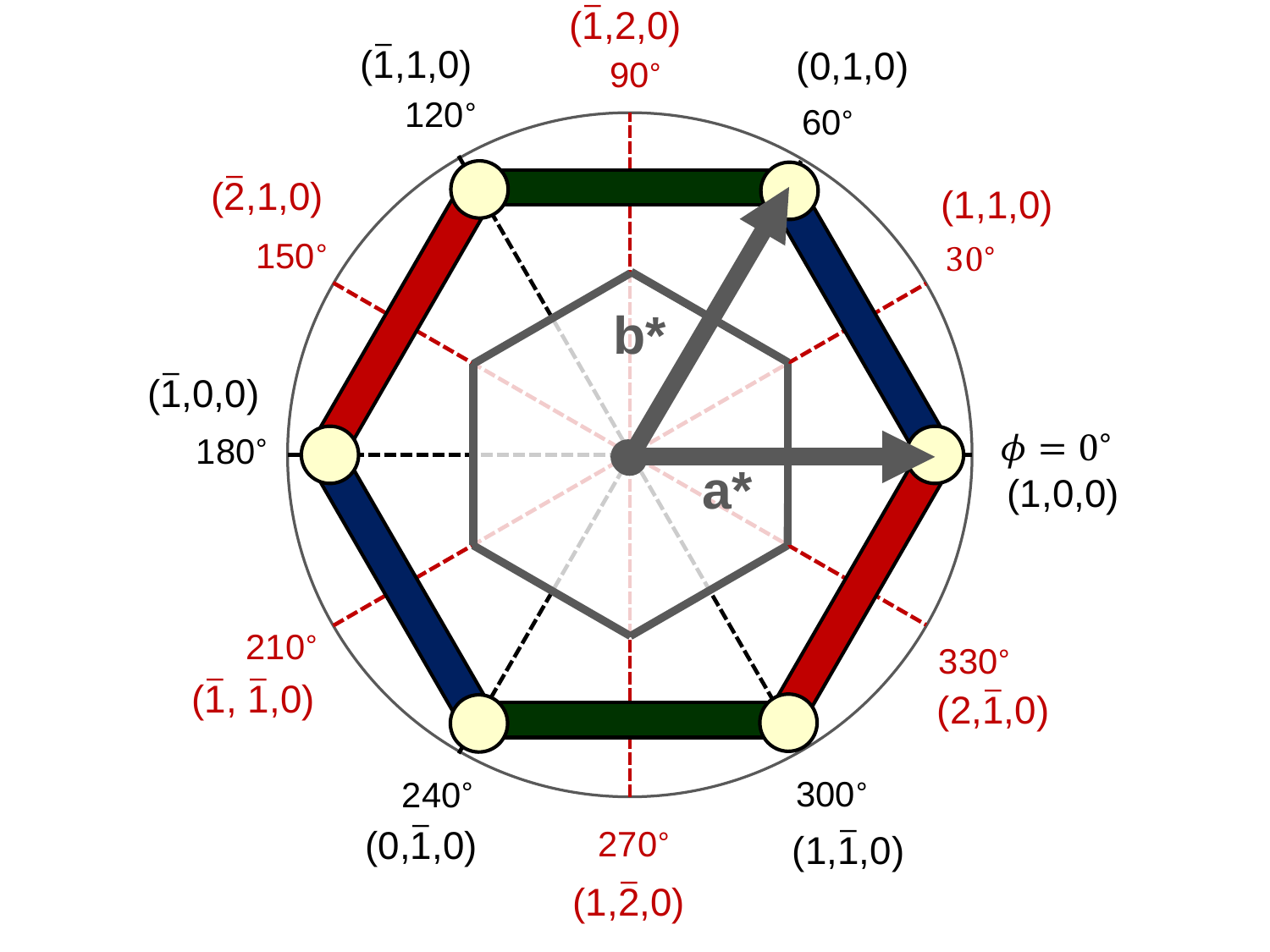}
\caption{\textbf{Definition of the in-plane angle \boldmath$\phi$} within the first Brillouin zone of the 2D reciprocal lattice (grey hexagon). Vectors which are symmetry-equivalent to (1,0,0) (black) and (1,1,0) (red) occur at angles of $\phi \equiv 0^\circ \mod 60^\circ$ and $\phi \equiv 30^\circ \mod 60^\circ$, respectively. The corresponding real-space orientation of the Ru-Ru bonds (red/blue/green hexagon) is overlaid. 
}
\label{fig:honeycomb}
\end{figure}

Quantum spin liquids constitute a most fascinating class of many-body states whose emergent excitations are fundamentally different from that of conventional magnets. The Kitaev model on the honeycomb lattice \cite{Kit06} is a unique and solvable example: Local spin flips fractionalize into itinerant Majorana fermions and Ising gauge-flux excitations.

The search for experimental realizations of the Kitaev model has uncovered a number of insulating honeycomb-lattice magnets where strong spin-orbit coupling generates $J=1/2$ local moments subject to bond-dependent Ising interactions \cite{Jac09,Cha10}. Most prominent are \aio\ ($A = \mathrm{Na}, \mathrm{Li}$) and {\rucl}; however, antiferromagnetic long-range order is realized at low temperatures in all of these materials.
Among them, {\rucl} has attracted immense attention \cite{plumb14,sears15,banerjee16} for two reasons:
(i) Spectroscopic experiments have detected \cite{banerjee16,banerjee17} clear signatures of fractionalized excitations over a significant range of energies which have been interpreted in terms of proximate spin liquid behavior \cite{banerjee16,banerjee17,gohlke17}.
(ii) A magnetic field applied in the honeycomb plane quickly suppresses magnetic order in favor of a quantum disordered state whose properties have been controversially debated\cite{wolter17,sears17,baek17,zheng17,leahy17,winter18,hentrich18}. Some recent experimental results hint at multiple field-induced phases as the zigzag phase is suppressed: preliminary ac susceptibility showed evidence for two transitions \cite{banerjee2018}, and a new report of quantized thermal Hall conductivity suggests the existence of a topological phase transition at a field above suppression of the zigzag ordered state \cite{Matsuda18}.
On the theoretical front, a debate has revolved around the most appropriate microscopic Hamiltonian to describe the magnetism of {\rucl} (refs.~\onlinecite{Kimchi11,Cha13,rau14,perkins14,ioannis15,winter16,winter17,janssen17}), with progress only made very recently.
However, a proper interpretation of the magnetic field effects and the nature of the proximate spin liquid are important open issues which require clarification if progress towards realizing a Kitaev spin liquid is to be made.

\begin{figure*}[t]
\includegraphics[width=0.95\textwidth]{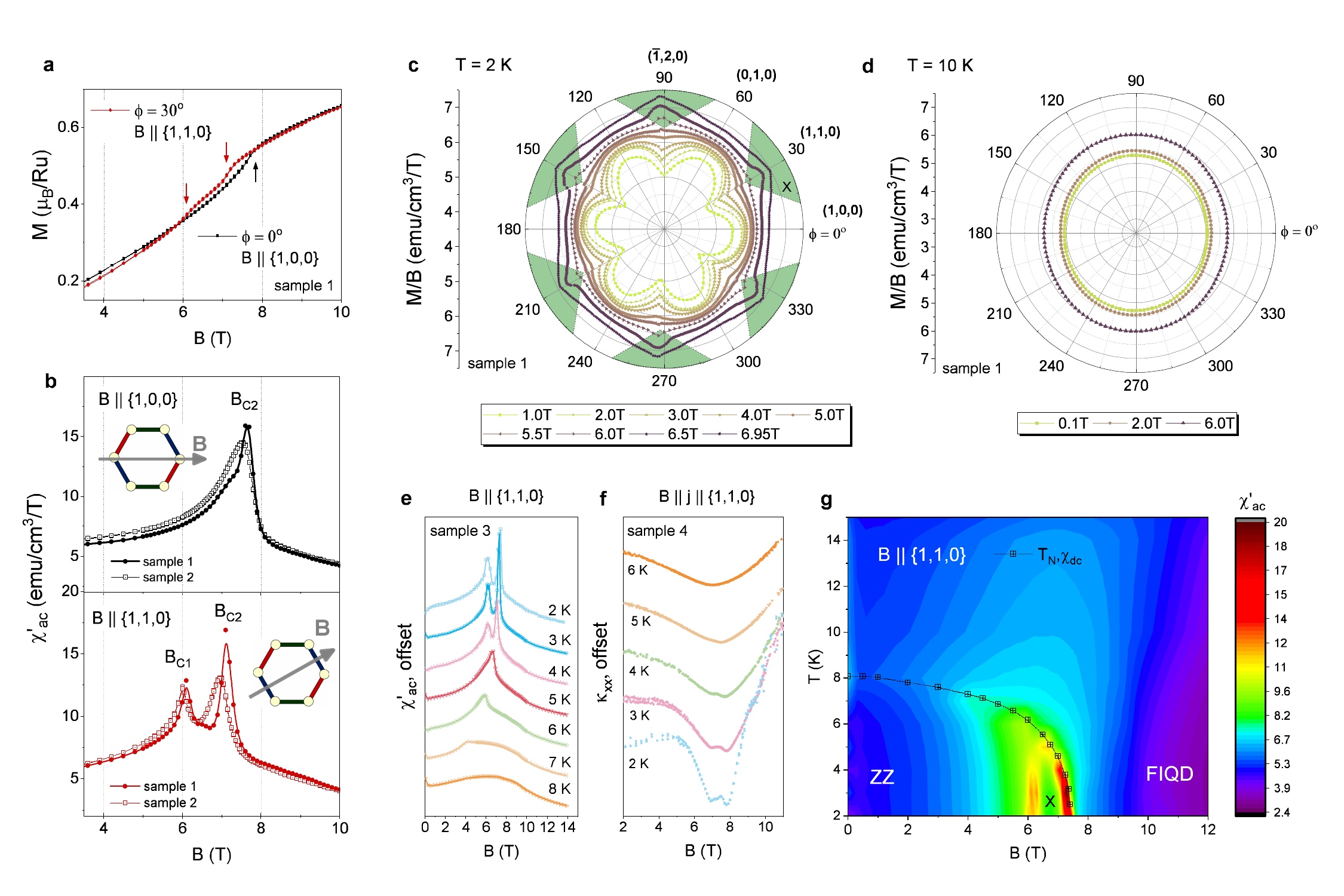}
\caption{
\textbf{Experimental evidence for a field-induced intermediate phase.}
(a) Field-dependent magnetization for different in-plane field directions $\phi$ at $T=2$~K. Arrows indicate kinks in the magnetization. 
(b) Real part of ac susceptibility $\cac$ as a function of dc magnetic field for in-plane field directions $\mathbf B\parallel\{1,0,0\}$ ($\phi \equiv 0^\circ \mod 60^\circ$) (upper panel) and  $\mathbf B\parallel\{1,1,0\}$ ($\phi \equiv 30^\circ \mod 60^\circ$) (lower panel), at $T=2$~K with two peaks indicating two phase transitions as function of field strength. The frequency of the 1~mT ac exciting field is 1~kHz.
(c) Angle dependence of the dc magnetization for various field strengths, plotted as $M/B$, at $T=2$~K, showing sixfold oscillations as function of $\phi$ whose signs reverse around $6$~T. The angular width of the intermediate-field region, X, is illustrated in green.
(d) Angle dependence of the magnetization at $T=10$~K.
Measurements of (e) $\cac$ and (f) thermal conductivity $\kappa_{xx}$ at various fixed temperatures show two anomalies as a function of $\{1,1,0\}$ magnetic field strength for $T < 5$ K. Curves are offset for clarity. 
(g) Phase diagram of {\rucl} as function of temperature and field along $\{1,1,0\}$ ($\phi=30^\circ$), constructed from $\cac$ data in (e) (color-coded). The N\'eel temperature extracted from dc magnetization measurements \cite{banerjee2018} is shown as squares. The intermediate phase, indicated by ``X'', occurs between the zigzag (ZZ) and field-induced quantum disordered (FIQD) state.
}
\label{fig:exp}
\end{figure*}

\begin{figure*}[t]
\includegraphics[width=0.8\textwidth]{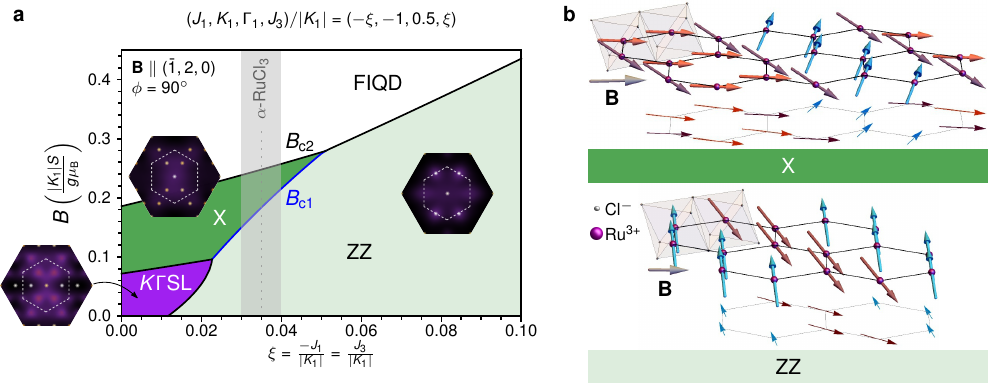}
\caption{
\textbf{Theoretical results for the classical Heisenberg-Kitaev-$\Gamma$ model.}
(a) Phase diagram as function of magnetic field $\mathbf B \parallel (\bar 1, 2, 0) \in \{1,1,0\}$ and strength of Heisenberg interactions $\xi = J_3/|K_1| = -J_1/|K_1|$ for fixed $\Gamma_1/|K_1|=0.5$ and $K_1 < 0$, showing the zigzag (ZZ) and field-induced quantum disordered (FIQD) phases, the spin liquid of the Kitaev-$\Gamma$ model ($K\Gamma$SL), and the novel field-induced phase (X). The vertical shaded region indicates the parameter regime relevant for {\rucl}.
Insets: static structure factors, showing the location of the Bragg peaks in the extended Brillouin zone.
For this direction, the field selects two domains of ZZ and X, respectively, indicated in both cases by the two Bragg peaks at finite wavevector in the first Brillouin zones (white dashed hexagons).
(b) Spin configurations at $\xi = 0.035$ of the X phase for $B = 0.22 |K_1 S|/(g \mu_\mathrm{B})$, with 6-site unit cell (top), %
and of the canted zigzag phase for $B = 0.10 |K_1 S|/(g \mu_\mathrm{B})$, with 4-site unit cell (bottom).
}
\label{fig:th}
\end{figure*}

Using multiple experimental probes, we report the discovery of a novel field-induced phase, and map its phase diagram as a function of magnetic field direction and temperature, 
several key features of which can be understood within our theoretical description of an intermediate ordered state. This allows us to reveal the nature of the exotic collective quantum state that is proximate to \rucl\ at zero field~\cite{banerjee16,banerjee17}: a spin liquid whose properties are distinct from the pure Kitaev solution.


\paragraph*{Experimental results.} 
When a magnetic field $\mathbf B$ is applied parallel to a Ru-Ru bond [corresponding to one of the symmetry-equivalent $(1,0,0)$, $(0,1,0)$, or $(-1,1,0)$ directions, see Fig.~\ref{fig:honeycomb}], the magnetization at 2~K shows a single kink at $\simeq  7.6$~T (with minor variation from one sample to another, see e.g. Fig.~\ref{fig:exp}b) in the vicinity of the well-documented field-induced suppression of the zigzag ordered phase \cite{wolter17,sears17,baek17,zheng17,leahy17,winter18,hentrich18} (Fig.~\ref{fig:exp}a). However, magnetic field directed perpendicular to a bond along a $\{1,1,0\}$-equivalent direction reveals a second feature near 6~T, well below the purported transition into the field-induced disordered phase. The anisotropy of the critical fields within the honeycomb plane is clearly visible in ac susceptibility $\cac$ measurements (Fig.~\ref{fig:exp}b). Two well-separated anomalies in $\cac$ at $B_\mathrm{c1} \simeq 6$~T and $B_\mathrm{c2} \simeq$ 7--7.3~T as a function of $\{1,1,0\}$ field strength converge and shift slightly higher to $B_\mathrm{c2} \simeq 7.6$~T in a $\{1,0,0\}$ field. This behavior repeats every 60$^{\circ}$, consistent with the symmetry of the honeycomb lattice, and has been reproduced in a number of samples.

Fig.~\ref{fig:exp}c shows the angle-dependence of the magnetization obtained via a sample rotation stage up to a maximum field of 7~T. Here, $\phi$ is the angle between $\mathbf{a^*}$ and $\mathbf{B}$ (see Fig.~\ref{fig:honeycomb}). At moderate fields $\gtrsim 1$~T, angle-resolved magnetization below $\TN=7$~K exhibits a six-fold symmetry with maxima at $\phi \equiv 0^\circ \mod 60^\circ$, where the field coincides with a bond-parallel $\{1,0,0\}$ direction. The amplitude of this oscillation decreases with increasing field. At 6~T, a distinct set of maxima appear in a narrow range of $\phi$ around the $\{1,1,0\}$ directions $\phi \equiv 30^\circ \mod 60^\circ$. The onset of the new oscillation coincides with the lower anomaly $B_\mathrm{c1}$ in Fig.~\ref{fig:exp}a,b, suggesting that for fixed fields between 6 and 7~T, the system alternates between the zigzag phase and the intermediate-field phase defined by $B_\mathrm{c1} < B < B_\mathrm{c2}$ as a function of $\phi$.

The double-peak behavior in $\cac (B)$ in a $\{1,1,0\}$ magnetic field emerges several Kelvin below the N\'eel transition, becoming distinct only for $T\leq 4$~K (Fig.~\ref{fig:exp}e). Thermal conductivity measurements (Fig.~\ref{fig:exp}f) exhibit consistent behavior. A minimum in $\kappa_{xx} (B)$ marking the critical field for the suppression of the zigzag order in RuCl$_{3}$ has been previously reported \cite{leahy17,hentrich18,Yu18}. With $\mathbf B \parallel\{1,1,0\}$, we find that this feature splits into two distinct minima below 5~K. We note that the $\kappa_{xx} (B)$ minima near 7 and 7.8~T are larger than the analogous critical fields in susceptibility data; the detailed field-dependence of magnetic contributions to phonon scattering and $\kappa_{xx}$ enhancement across the two transitions are not well understood and likely play a role in the discrepancy. The boundaries of the intermediate-field region (denoted by ``X'') as a function of temperature and $\{1,1,0\}$ field strength are shown in a $\cac$ intensity plot in Fig.~\ref{fig:exp}g.
The emergence of this distinct region below $T_\mathrm{N}$ indicates the existence of a novel magnetically ordered phase.


\paragraph*{Theoretical results.}

We employ a Heisenberg-Kitaev-$\Gamma$ spin model with a dimensionless parameter $\xi \geq 0$, where $\xi$ measures the strength of the isotropic Heisenberg interaction relative to the bond-dependent interactions, see Methods for details of the Hamiltonian. This model is inspired by \emph{ab initio} calculations~\cite{winter16} and has recently been shown to reproduce various measurements in \rucl\ both at zero~\cite{winter17} and finite~\cite{janssen17, winter18} field.
The classical phase diagram in an in-plane magnetic field $\mathbf B \parallel (\bar 1, 2, 0)$ is displayed in Fig.~\ref{fig:th}a. For $\xi > 0.012$, the low-field ground state has a zigzag order (``ZZ'' in Fig.~\ref{fig:th}) with a 4-site magnetic unit cell and ordering wavevector $\mathbf Q = \mathbf{M}$, where $\bf M$ denotes the center of an edge of the first Brillouin zone, as observed experimentally in \rucl\ (refs.~\onlinecite{sears15,banerjee2018}).
At elevated fields and $0.024 < \xi < 0.051$, a novel ordered state is stabilized (``X'' in Fig.~\ref{fig:th}). This field-induced intermediate phase is characterized by a 6-site magnetic unit cell consisting of three inequivalent pairs of parallel spins, resulting in a noncoplanar spin structure. The real-space spin configuration, Fig.~\ref{fig:th}b, can be understood as a period-3 pattern of a ferromagnetic zigzag chain alternating with two non-collinear zigzag chains. It consequently shows Bragg peaks at $\mathbf Q = \frac{2}{3}\mathbf{M}$.
At zero field, this is a metastable phase that is characterized by a finite total moment, i.e., it is \emph{ferrimagnetic}, which naturally explains its stabilization in a finite field.
For very small $\xi$, the quantum ground state is a spin liquid studied in refs.~\onlinecite{catun18,gohlke18}. The classical analogue of this novel Kitaev-$\Gamma$ spin liquid is characterized by a large ground-state degeneracy at zero field and covers a finite region in the $B$-$\xi$ phase diagram (``$K\Gamma$SL'' in Fig.~\ref{fig:th}a).
%

\begin{figure*}[t]
\includegraphics[width=0.7\textwidth]{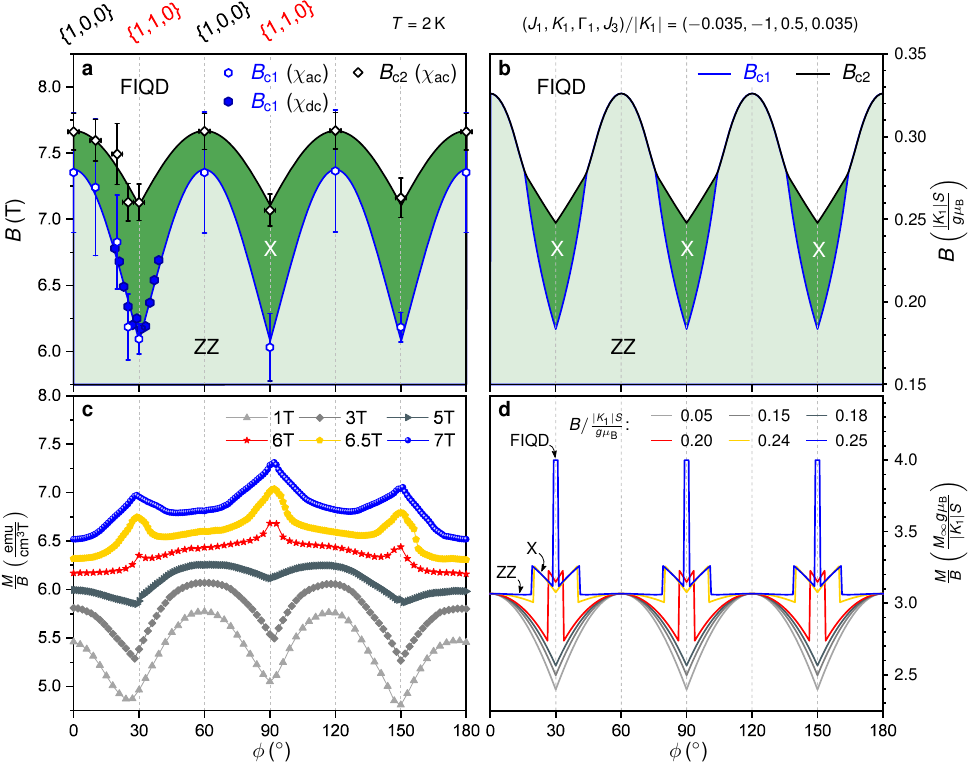}
\caption{
\textbf{Comparison experiment--theory.}
(a,b): Phase diagram as function of field strength $B$ and angle $\phi$ from (a) ac and dc susceptibility measurements on {\rucl} at $T=2$~K and (b) the classical Heisenberg-Kitaev-$\Gamma$ model. 
The solid lines in (a) are a guide to the eye.
(c,d): Longitudinal magnetization, plotted as $M/B$, as function of field angle $\phi$ for different $B$ of (c) {\rucl} at $T=2$~K and (d) the classical Heisenberg-Kitaev-$\Gamma$ model for $T=0$.
In (d) the peaks in the high-field curves (blue) at $\phi = 30^\circ \mod 60^\circ$ arise from the transition to the FIQD phase at $B_\mathrm{c2}$, while the local maxima on either side of the peaks occur at $B_\mathrm{c1}$.
}
\label{fig:comp}
\end{figure*}


\paragraph*{Comparison experiment--theory.}
Fig.~\ref{fig:comp} shows the comparison of the phase diagram (a,b) and longitudinal magnetization (c,d) between \rucl\ (left) and the Heisenberg-Kitaev-$\Gamma$ model with $\xi = 0.035$ (right). 
Key characteristic features of the experiment are reproduced by the model calculation:
(i)~Both the critical fields and the magnetization curves have a 60$^\circ$ periodicity, consistent with the $C_3$ rotational symmetry of the honeycomb lattice. (We note that the symmetry is approximate in the experimental magnetization; see Supplementary Information for a discussion of domain effects.)
(ii)~The critical fields are maximal for a field along a Ru-Ru bond ($\phi \equiv 0^\circ \mod 60^\circ$) and minimal for a field perpendicular to a bond ($\phi \equiv 30^\circ \mod 60^\circ$).
(iii)~The width of the intermediate phase is maximal when the critical field is minimal, and vice versa. (We note, however, that the intermediate phase vanishes near $\phi \equiv 0^\circ \mod 60^\circ$ for the parameter set used in Fig.~\ref{fig:comp} in the classical model, while it may have a finite width in \rucl\ for these angles, see Supplementary Information for details of the critical field determination from the $\cac$ data.)
(iv)~The magnetization $M$ for fields below the minimal lower critical field $B_\mathrm{c1}(30^\circ)$ is maximal for $\phi \equiv 0^\circ \mod 60^\circ$ and has kink-like minima at $\phi \equiv 30^\circ \mod 60^\circ$. (We note that $M/B$ does not increase with $B$ for $\phi = 0^\circ \mod 60^\circ$ in the calculation, in contrast to the experimental curve. This is a known anomaly of the classical approximation~\cite{janssen17} that will resolve upon the inclusion of quantum fluctuations.)
(v)~In the intermediate regime, $B_\mathrm{c1}(30^\circ)< B < B_\mathrm{c1}(0^\circ)$, on the other hand, the magnetization becomes maximal for $\phi \equiv 30^\circ \mod 60^\circ$.
(vi)~Finally, taking the ab-initio guided values~\cite{winter17} for the magnetic interaction scale $K_1 \simeq -5~\text{meV}$ and the in-plane $g$-factor $g_{ab} \simeq 2$, we find $B_\mathrm{c1}(0^\circ) \simeq 7~\text{T}$, in remarkable agreement with the experimental critical field.
%


\paragraph*{Discussion.}
We have identified a novel field-induced intermediate phase in the Kitaev material {\rucl}. Our comprehensive modelling establishes a non-coplanar phase with 6-site magnetic unit cell that captures the experimentally observed features of the intermediate-field region. 
A recent work\cite{Matsuda18} reported quantization of the thermal Hall conductivity in tilted fields with in-plane components $7~\text{T} \lesssim B_\parallel \lesssim 9~\text{T}$, implying Majorana edge currents above the suppression of the zigzag phase at 7~T and a transition to a non-topological phase at 9~T.
It would be interesting to search for such a topological transition within a full quantum simulation of our Heisenberg-Kitaev-$\Gamma$ model.
The novel intermediate phase we have found, however, occurs at smaller field strengths and below $T_\mathrm{N}$, indicating long-range magnetic order that can be understood within a semiclassical approximation.
We note that our susceptibility data appear featureless above $B_{\mathrm{c}2}$ in most samples checked, however, some additional small kinks were observed between 8 and 10 T in a sample with mixed $\sim$7 and $\sim$14~K N\'eel transitions (see Supplementary Information).

Most importantly, our intermediate ordered phase is the leading field-induced instability of the quantum spin liquid of the Kitaev-$\Gamma$ model\cite{catun18}. Its appearance implies that \rucl\ is proximate to the Kitaev-$\Gamma$ spin liquid.
This Kitaev-$\Gamma$ spin liquid reacts sensitively to small bond anisotropies induced by a monoclinic distortion, as present in \rucl\ (ref.~\onlinecite{plk2018}), by opening up a gap in the spinon spectrum\cite{gohlke18}. This is in contrast to the pure Kitaev model, which remains gapless in the presence of such bond anisotropies\cite{Kit06}.
The spin-liquid signatures observed in dynamical probes of \rucl\ (refs.~\onlinecite{banerjee16, banerjee17}) should therefore be associated with a proximity to a topologically ordered Kitaev-$\Gamma$ spin liquid, with a spinon excitation gap that may be expected to be of the order of the effective size of the bond anisotropies~\cite{gohlke18}.
Whether this low-field Kitaev-$\Gamma$ spin liquid is related to the field-induced quantum disordered state observed above $B_\mathrm{c2}$ is an interesting question for future investigations.


\section*{Methods}
\paragraph*{Experimental.}
Susceptibility measurements were performed in $\alpha$-RuCl$_3$ single crystals prepared by a vapor transport method \cite{banerjee17} and oriented by Laue using the trigonal reciprocal setting (see definition in Fig.~\ref{fig:honeycomb}). Angle-resolved dc magnetization measurements were collected using a sample rotation stage in a 7~T SQUID magnetometer. dc magnetization, ac susceptibility, and thermal transport  measurements were performed up to 14~T at various fixed angles in a Physical Property Measurement System (Quantum Design). 

\paragraph*{Theoretical.}
The theoretical results have been obtained by employing the bond-dependent spin Hamiltonian
\begin{align}
\label{eq:H}
\mathcal{H} & =
\sum_{\langle ij\rangle_\gamma} \left[
J_1 \mathbf S_i \cdot \mathbf S_j  + K_1 S_i^\gamma S_j^\gamma
+ \Gamma_1 \left( S_i^\alpha S_j^\beta + S_i^\beta S_j^\alpha\right)
\right]
\nonumber \\
&\quad + \sum_{\lllangle ij\rrrangle}
J_3 \mathbf S_i \cdot \mathbf S_j
- \mu_\mathrm{B}\mathbf B \cdot g \sum_{i} \mathbf S_i
\end{align}
with Heisenberg, Kitaev, and symmetric off-diagonal exchange terms on nearest-neighbor bonds,
Heisenberg exchange on third-neighbor bonds, and a uniform in-plane magnetic field. 
We have assumed a diagonal $g$-tensor, 
$g = 
\left(\begin{smallmatrix}
g_{ab} & 0 & 0\\
0 & g_{ab} & 0\\
0 & 0 & g_{c^*}\\
\end{smallmatrix}\right)$
in the crystallographic $(\mathbf a, \mathbf b, \mathbf c^*)$ basis,
with an isotropic in-plane $g_a = g_b \equiv g_{ab}$, in agreement with previous results\cite{agrestini2017}.

We have used $(J_1, K_1, \Gamma_1, J_3) = A ( - \xi, -1, 0.5, \xi)$, where $A$ sets the overall energy scale and $\xi\geq 0$ parametrizes the deviation from the Kitaev-$\Gamma$ limit. Previously, this model with $A=5~\mathrm{meV}$ and $\xi = 0.1$ has been suggested as an effective description for \rucl\ (refs.~\onlinecite{winter17, janssen17, winter18}). Fig.~\ref{fig:th}a shows that a slight modification of this parameter choice fully accounts for the existence of the novel field-induced ordered phase and the proximity to a spin-liquid regime.

To identify the zero-temperature ground state of this model in finite field, we have utilized classical Monte Carlo (MC) simulations, combining single-site and parallel-tempering updates in order to equilibrate the spin configurations at the lowest temperatures. From the MC data, we have computed the static spin structure factor
\begin{align}
S(\mathbf k) = \frac{1}{N} \sum_{i,j=1}^N \left\langle \mathbf S_i \cdot \mathbf S_j \right\rangle \mathrm{e}^{\mathrm{i} \mathbf{k} \cdot (\mathbf R_i - \mathbf R_j)}\,,
\end{align}
shown in the insets of Fig.~\ref{fig:th}(a). Here, $\mathbf R_i$ denotes the lattice vector at site $i$ and $N$ the number of sites.
The MC simulations are supplemented with an analytical parametrization of the classical phases, allowing us to compute the full zero-temperature phase diagrams shown in Fig.~\ref{fig:th}(a) and Fig.~\ref{fig:comp}(b) and the magnetization curves in Fig.~\ref{fig:comp}(d) in a numerically exact way. The analytical parametrization also reveals the real-space spin structure of the field-induced intermediate state, as shown in Fig.~\ref{fig:th}(b), as well as ferrimagnetic nature of this (then metastable) state at zero field. We have explicitly verified for various field angles $\phi$ and values of the parameter $\xi$ that the ground-state energy $\varepsilon$ obtained by employing the analytical parametrization agrees with the MC results within a precision of $\Delta \varepsilon / \varepsilon \lesssim 10^{-8}$, see Supplementary Information for details.

\paragraph*{Data availability.}
The data that support the plots within this paper and other findings of this study are available from P.L.K.\ (experimental) and L.J.\ (theoretical) upon request.


\begin{acknowledgments}
We thank B.\ B\"uchner, R.\ Valent{\'i}, and A.\ U.\ B.\ Wolter for enlightening discussions. 
The computations were partially performed at the Center for Information Services and High Performance Computing (ZIH) at TU Dresden.
This research was supported by the DFG through SFB 1143 (L.J.\ and M.V.) and GRK 1621 (M.V.).
E.C.A.\ was supported by FAPESP (Brazil) Grant No.\ 2013/00681-8 and CNPq (Brazil) Grant No.\ 302065/2016-4.
P.L.K.\ and D.G.M.\ acknowledge support from Gordon and Betty Moore Foundation's EPiQS Initiative through Grant GBMF4416.
Work at ORNL was supported by the US DOE, Office of Science, Basic Energy Sciences, (J.Q.Y.)~Materials Sciences and Engineering Division, and (C.B.\ and S.E.N.) Division of Scientific User facilities, under contract number DE-AC05-00OR22725. 
\end{acknowledgments}


\section*{Author contributions}
P.L.K., L.J., S.R., D.G.M., and M.V.\ conceived the project. J.Q.Y.\ and P.L.K.\ made the single crystals and C.B.\ conducted the Laue measurements. P.L.K.\ performed the susceptibility and magnetization measurements. J.Q.Y. and P.L.K.\ performed the heat transport measurements. P.L.K., C.B., D.G.M., and S.E.N.\ analyzed the experimental data. L.J.\ and M.V.\ conceived the model. L.J.\ and E.C.A.\ carried out the analytical and numerical computations and analyzed the theoretical data. P.L.K., L.J., and M.V.\ prepared the first draft and all authors contributed to the writing of the manuscript.


\section*{Additional information}
Supplementary information accompanies this paper. 
Correspondence and requests for materials should be addressed to P.L.K.\ and L.J.
The authors declare no competing interests.


\hypersetup{
    colorlinks,
    linkcolor={blue},
    citecolor={blue},
    urlcolor={blue}
}

\renewcommand{\theequation}{S\arabic{equation}}
\renewcommand{\thefigure}{S\arabic{figure}}
\renewcommand{\thetable}{S\arabic{table}}

\graphicspath{{./}{./plots/}}



\clearpage

\onecolumngrid

\renewcommand{\thefigure}{S\arabic{figure}}

\setcounter{figure}{0}

\renewcommand{\thetable}{S\arabic{table}}

\setcounter{table}{0}

\section*{Supplementary Information:\\
Field-induced intermediate phase in \rucl: \\
Non-coplanar order, phase diagram, and proximate spin liquid}

\author{P. Lampen-Kelley\cite{contrib, plk}}
\affiliation{Department of Materials Science and Engineering, University of Tennessee, Knoxville, TN 37996, U.S.A.}
\affiliation{Materials Science and Technology Division, Oak Ridge National Laboratory, Oak Ridge, TN 37831, U.S.A.}

\author{L. Janssen\cite{contrib, lj}}
\affiliation{Institut f\"ur Theoretische Physik, Technische Universit\"at Dresden,
01062 Dresden, Germany}

\author{E. C. Andrade}
\affiliation{Instituto de F\'{i}sica de S\~ao Carlos, Universidade de S\~ao Paulo, C.P. 369,
S\~ao Carlos, SP 13560-970, Brazil}

\author{S. Rachel}
\affiliation{School of Physics, University of Melbourne, Parkville, VIC 3010, Australia}

\author{J.-Q.~Yan}
\affiliation{Materials Science and Technology Division, Oak Ridge National Laboratory, Oak Ridge, TN 37831, U.S.A.}

\author{C. Balz}
\affiliation{Neutron Scattering Division, Oak Ridge National Laboratory, Oak Ridge, TN 37831, U.S.A.}

\author{D. G. Mandrus}
\affiliation{Department of Materials Science and Engineering, University of Tennessee, Knoxville, TN 37996, U.S.A.}
\affiliation{Materials Science and Technology Division, Oak Ridge National Laboratory, Oak Ridge, TN 37831, U.S.A.}

\author{S. E. Nagler}
\affiliation{Neutron Scattering Division, Oak Ridge National Laboratory, Oak Ridge, TN 37831, U.S.A.}

\author{M. Vojta}
\affiliation{Institut f\"ur Theoretische Physik, Technische Universit\"at Dresden,
01062 Dresden, Germany}


\maketitle
\twocolumngrid


\begin{figure}[t]
\includegraphics[width=0.4\textwidth]{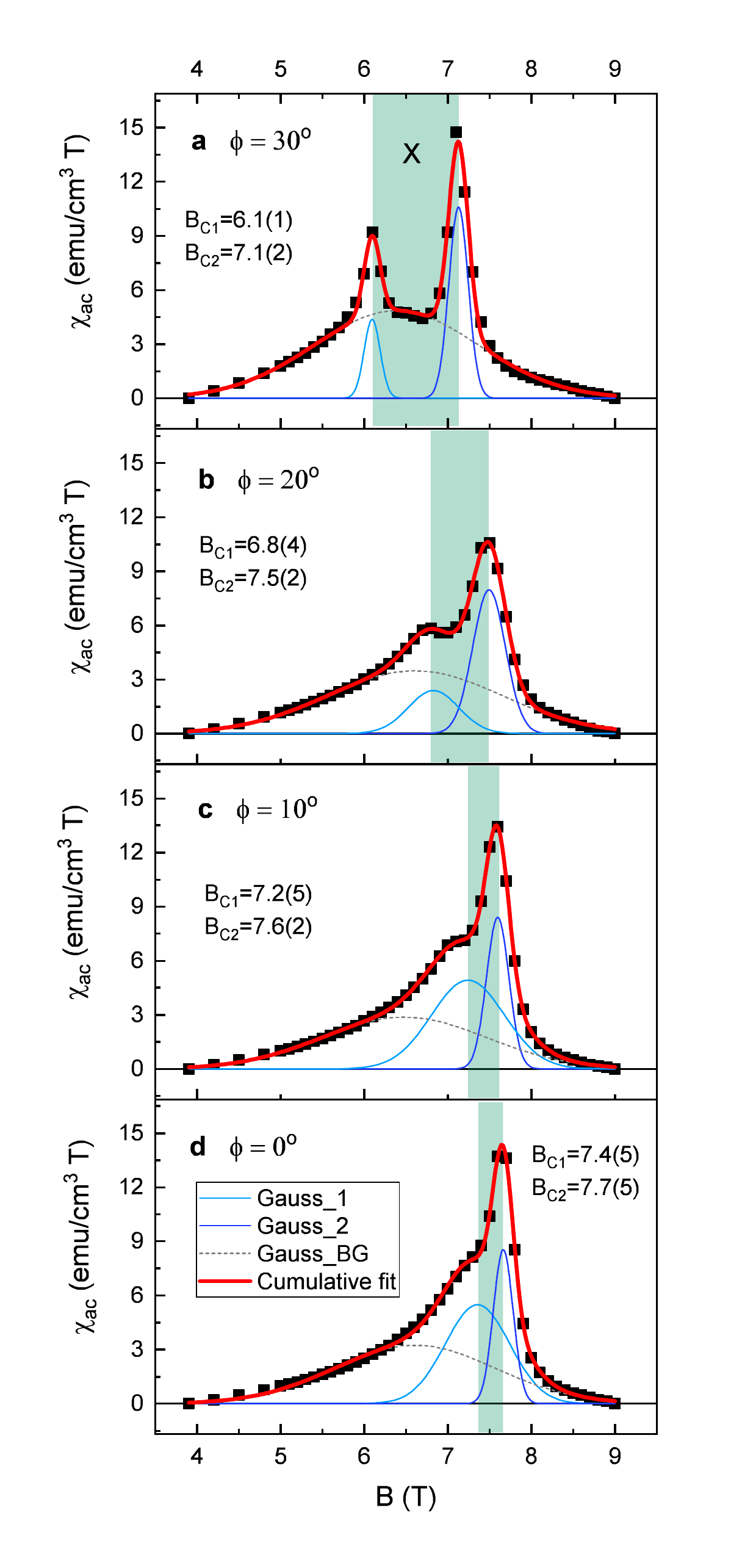}
\caption{
\textbf{Intermediate phase width for various angles}
from ac susceptibility curves as a function of dc magnetic field for fixed angles of (a) $\phi=30^{\circ}$, (b) $\phi=20^{\circ}$, (c) $\phi=10^{\circ}$, and (d) $\phi=0^{\circ}$ in Sample 1 at $T = 2$ K. The ac exciting field was 1 mT in amplitude with a frequency of 1 kHz. A linear background is subtracted from the data shown between 4 and 9 T. Three Gaussian peaks are used to model the anomalies at the critical fields $B_{\mathrm{c}1}$ and $B_{\mathrm{c}2}$ and the background. 
}
\label{fig:Xfit}
\end{figure}

\section{Determination of the critical fields from ac susceptibility data}

The angle dependence of the critical fields $B_{\mathrm{c}1}$ and $B_{\mathrm{c}2}$ shown in Fig.~4a of the main manuscript were determined from ac susceptibility curves collected at various fixed angles at $T = 2$ K for Sample 1. Fig.~\ref{fig:Xfit} shows the data for $\phi = 0^{\circ}$, $10^{\circ}$, $20^{\circ}$, and $30^{\circ}$ between 4 and 9 T after subtraction of a linear background. The two anomalies representing $B_{\mathrm{c}1}$ and $B_{\mathrm{c}2}$ were modeled as Gaussian peaks; the overall increase in the susceptibility in the critical region was treated phenomenologically by introducing a third, broad Gaussian peak as a background on which the sharper anomalies are superimposed. The width of the intermediate phase defined as the field interval between the two peak centers narrows from 1 T at $\phi=30^{\circ}$ to 0.7~T at $\phi=20^{\circ}$ and 0.4 T at $\phi=10^{\circ}$. At $\phi=0^{\circ}$, a shoulder on the sharp peak at $B_{\mathrm{c}2}$ suggests that the intermediate still has a finite width of 0.3 T, in contrast to the model in Fig.~4b of the main manuscript, although we note that 0.3 T is within the error bars on the critical fields taken as one half of the fitted FWHM of the peaks. A distinct shoulder was not evident at $\phi=0^{\circ}$ for Sample 2 shown in Fig.~2b of the main manuscript, however this sample exhibited slightly broader peaks overall as compared with Sample 1.   


\begin{figure*}[p]
\includegraphics[width=0.8\textwidth]{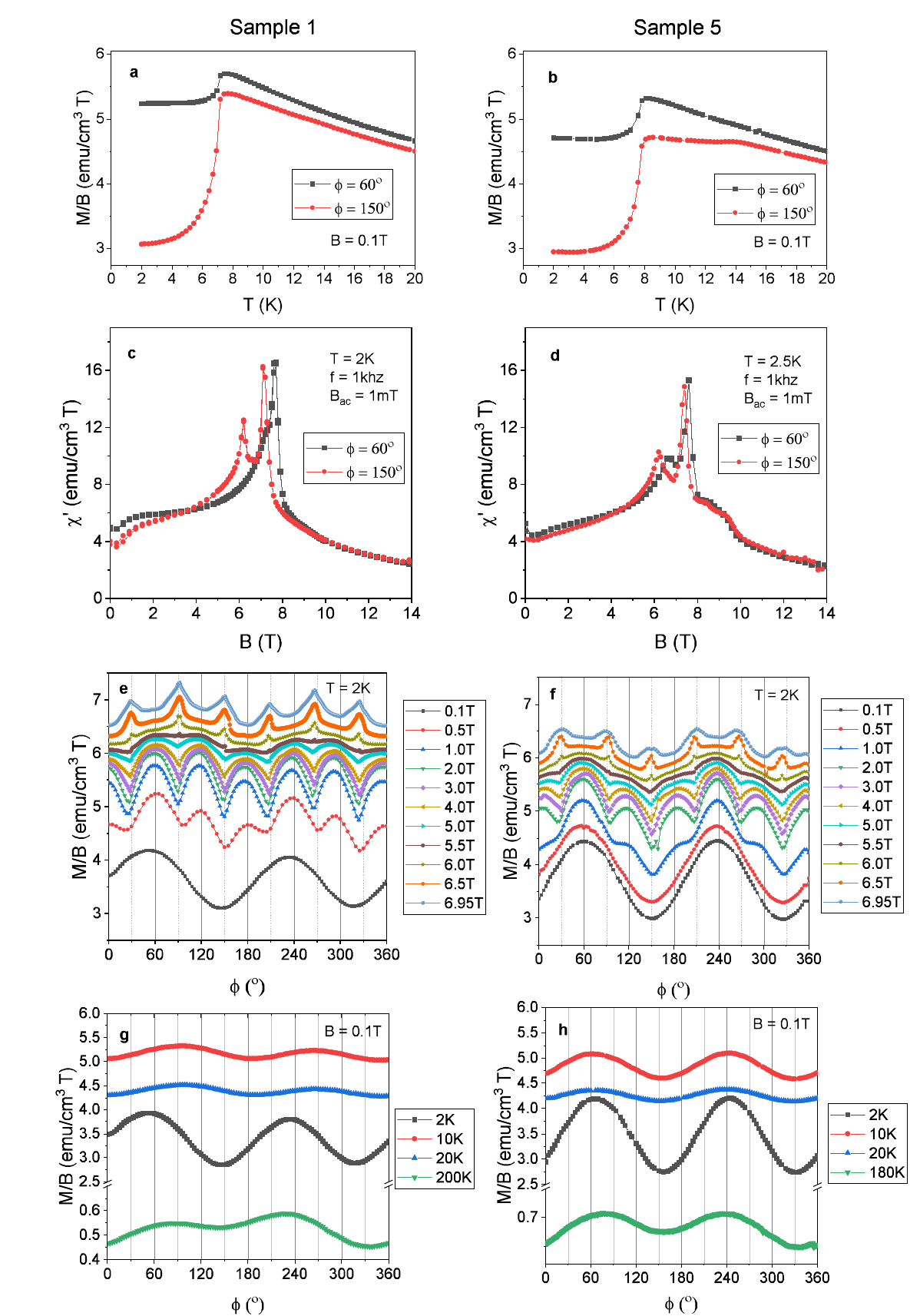}
\caption{
\textbf{Comparison of susceptibility measurements} in Sample 1 (a,c,e,g) with a single $T_\mathrm{N} \sim7$ K magnetic transition and Sample 5 (b,d,f,h) with an additional shoulder at $T_\mathrm{N2} \sim14$ K. 
(a,e) Temperature-dependent dc susceptibility at $\phi=60^\circ$ and $\phi=150^\circ$, corresponding to a \{1,0,0\}-type and \{1,1,0\}-type direction, respectively. (c,d) Field-dependent ac susceptibility at $T=2$ K for $\phi=60^\circ$ and $\phi=150^\circ$. (e,f) Angle-dependent dc susceptibility at various fixed magnetic fields for $T=2$ K. 
}
\label{fig:SampleCompare}
\end{figure*}

\section{Comparison of susceptibilities in single- and multi-transition crystals}

The samples for which data are presented in Fig.~2 of the main manuscript (Samples 1-4) showed no evidence of additional magnetic transitions above $\sim$7 K, which have been observed in samples that exhibit stacking disorder in the as-grown state \cite{Johnson2015}, or in which stacking faults are introduced in the course of extensive handling or deformation of the crystals \cite{Cao2016}. 
To compare with Sample 1, we performed a similar set of susceptibility measurements on another crystal ``Sample 5'' in which a small amount of the 14~K magnetic transition is visible in the susceptibility curve collected with field in the lower-magnetization \{1,1,0\}-type direction ($\phi=150^{\circ}$) (Fig.~\ref{fig:SampleCompare}b). The fixed-angle ac susceptibility curves collected up to 14 T in Sample 5 show anomalies at 6.2 and 7.3 T with $B_\mathrm{dc}$ along $\phi=150^{\circ}$, similar to the 6.0 and 7.3 T anomalies observed in Sample 1. However, Sample 5 exhibited additional smaller features at higher fields $\sim$8.5 and $\sim$9.5 T that are absent in Sample 1. In addition, we note that $B_{\mathrm{c}1}$ and $B_{\mathrm{c}2}$ shifted closer together and towards higher field in a \{1,0,0\}-type field ($\phi=60^{\circ}$) but remained separated in contrast to Sample 1.  

Angle-resolved dc magnetization measurements show similar overall features in the two samples at $T = 2$ K: a 2-fold oscillation at small fields consistent with unequally populated zigzag domains (see Sec.~\ref{sec:zz-domains}) and an additional 6-fold oscillation emerging at moderate fields with maxima at $\phi \equiv 0^\circ \mod 60^\circ$ and switching above $B_{\mathrm{c}1}\simeq 6$~T to exhibit maxima at $\phi \equiv 30^\circ \mod 60^\circ$ in the intermediate phase.  

In Sample 5 the 2-fold oscillation represented a more significant contribution to the total susceptibility than in Sample 1, and the phase of the 2-fold oscillation remained consistent at all fields and temperatures. In contrast, the 2-fold contribution in Sample 1 with maxima near $\phi=60^\circ$ and $240^\circ$ for $B=0.1$ T and $T=2$ K showed a 30$^\circ$ phase shift when increasing the magnetic field at $T = 2$ K producing global maxima at $\phi=90^\circ$ and $270^\circ$; the observation of the same phase shift upon warming above $T_\mathrm{N}$ at $B=0.1$ T likely indicates the presence of a weak underlying structural anisotropy that produces susceptibility maxima at $\phi=90^\circ$ and $270^\circ$. The different phase of the uniaxial anisotropy for 10 K $< T <$ 150 K between Sample 1 and Sample 5 suggests that it may be sensitive to structural domains below $T_\mathrm{S} \sim 150$ K. Above the transition to the monoclinic structure near 150 K, it appears that the phase in the 2-fold oscillation for Sample 1 once again shifts towards maxima at $\phi=60^\circ$ and $\phi=240^\circ$, consistent with Sample 5 and the discussion in a previous work \cite{plk2018}.


\begin{figure*}[tbp]
\includegraphics[width=\textwidth]{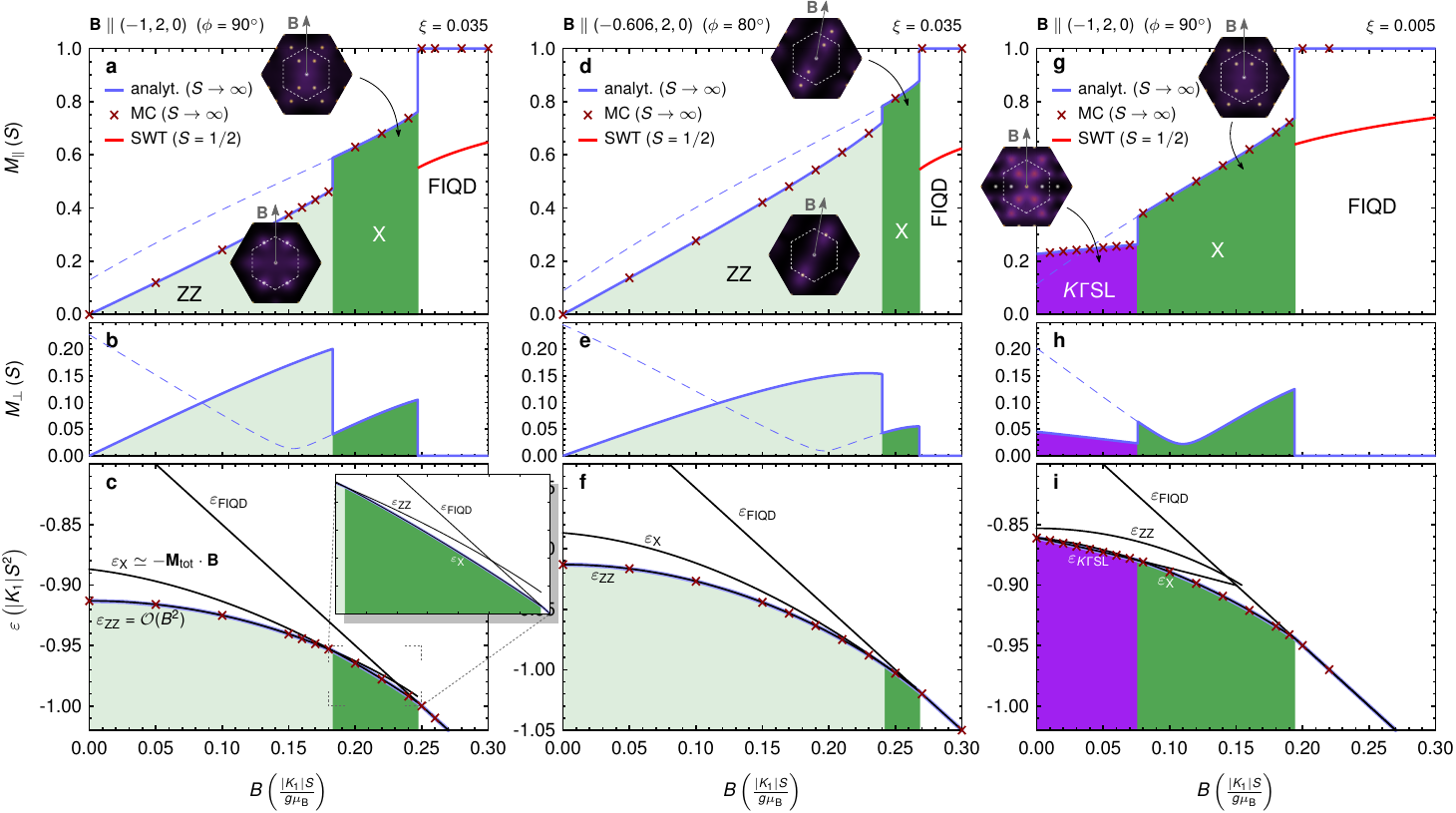}
\caption{
\textbf{Magnetization processes in the Heisenberg-Kitaev-$\Gamma$ model} for $(J_1, K_1, \Gamma_1, J_3)/|K_1| = (\xi, -1, 0.5, \xi)$.
(a,b,c): Longitudinal magnetization along field axis (a), transversal magnetization (b), and energy levels~(c) as a function of magnetic field $B$ along the $(-1,2,0) \in \{1,1,0\}$ direction ($\phi = 90^\circ$) for $\xi = -J_1/|K_1| = J_3/|K_1| = 0.035$. Dashed curve: magnetization of the metastable X state at small fields, indicating its ferrimagnetic character. Red curve: magnetization in FIQD phase for $S=1/2$. Insets: static structure factors for zigzag (ZZ) and X phases in the extended Brillouin zone, with the white dashed hexagons indicating the first Brillouin zone, which comprises two Bragg peaks at finite wavevectors as a result of the degeneracy between two ZZ and X domains, respectively, for this field direction (a). The transitions between ZZ and X and between X and FIQD, respectively, are first order due to level crossings (c).
(d,e,f): Same as (a,b,c), but for $\phi = 80^\circ$, for which the magnetic field favors single domains of ZZ and X states, as indicated by the structure factors~(d).
(g,h,i): Same as (a,b,c), but for small $\xi = 0.005$, close to the pure $K_1$-$\Gamma_1$ limit. Here, the zero-field ground state ($K\Gamma$SL) is highly degenerate and includes ferrimagnetic states as well. At some critical field, a first-order transition towards the intermediate X phase occurs.
}
\label{fig:mXB}
\end{figure*}

\section{Magnetization processes and structure factors in the $J_1$-$K_1$-$\Gamma_1$-$J_3$ model}

Fig.~\ref{fig:mXB} shows the field-dependent longitudinal magnetization $M_\parallel$ and transversal magnetization $M_\perp$, together with the classical energy of the different states for two different directions of the external field in the Heisenberg-Kitaev-$\Gamma$ model. These magnetization curves have been obtained for $(J_1, K_1, \Gamma_1, J_3) = A (-\xi, -1, 0.5, \xi)$ in the classical limit, $S \to \infty$, for $\xi = 0.035$ (Fig.~\ref{fig:mXB}a--f) and $\xi = 0.005$ (Fig.~\ref{fig:mXB}g,h,i), respectively. For comparison, we have also computed the quantum corrections to the longitudinal magnetization in linear spin wave theory for $S=1/2$ in the field-induced quantum disordered (FIQD) phase (red curves in Fig.~\ref{fig:mXB}a,d,g), demonstrating the strong influence of quantum fluctuations in this state.
At small fields, the (then metastable) X state is characterized by a finite total magnetization (dashed curves in Fig.~\ref{fig:mXB}a,b,d,e,g,h), indicating its ferrimagnetic character.
The energy gain in field is therefore linear in $B$, while the antiferromagnetic zigzag state gains energy only after spin canting, resulting in a quadratic dependence on $B$ to the leading order, see Fig.~\ref{fig:mXB}c,f. This explains why the X state is favored at elevated fields along the $\{1,1,0\}$ directions for $\xi = 0.035$.

The field-induced transition in and out of the intermediate X phase is characterized by a jump in the magnetization and an abrupt change of the static structure factor (see insets in Fig.~\ref{fig:mXB}a,d), indicating a first-order transition. 
A single-domain zigzag state has a Bragg peak at ordering wavevector $\mathbf Q = \mathbf{M}_i$, where $\mathbf{M}_i$, $i\in \{1,2,3\}$, denotes one of the three inequivalent centers of the edges of the first Brillouin zone. Correspondingly, there are three inequivalent zigzag domains, out of which typically only one is selected in small fields for a generic field direction, cf.\ Sec.~\ref{sec:zz-domains}.
However, for the higher-symmetric $\{1,1,0\}$ directions, as, e.g., the one employed in Fig.~\ref{fig:mXB}a, two of the three domains are related by symmetry (neglecting a possible bond anisotropy), and the Monte-Carlo-averaged low-temperature structure factor shown in the inset of Fig.~\ref{fig:mXB}a consequently displays two Bragg peaks at two out of the three $\mathbf{M}_i$ points. 
Tilting the field slightly away from this axis, by contrast, singles out the domain for which the angle between $\mathbf Q$ and $\mathbf B$ is minimized, see inset of Fig.~\ref{fig:mXB}d. 
This is consistent with the available neutron diffraction measurements in \rucl\ in fields up to $2~\mathrm{T}$ (ref.~\onlinecite{banerjee2018}).
Similarly, a single-domain X state has a single (up to inversion) Bragg peak in the first Brillouin zone at ordering wavevector $\mathbf Q = \frac{2}{3} \mathbf{M}_i$, $i \in \{1,2,3\}$. Just as the zigzag state, the X state therefore exhibits three inequivalent domains, and an external field~$\mathbf B$ selects the domain(s) for which the angle between $\mathbf Q$ and $\mathbf B$ is minimized. The structure factor of the intermediate phase shown in the inset of Fig.~\ref{fig:mXB}a therefore shows two Bragg peaks at finite wavevector corresponding to the two domains selected in a $\{1, 1, 0\}$ field, while only one (nontrivial) Bragg peak is present for the tilted field direction shown in the inset of Fig.~\ref{fig:mXB}d.

If we move closer to the pure Kitaev-$\Gamma$ limit (Fig.~\ref{fig:mXB}g,h,i) upon decreasing $\xi$, the zigzag ground state at zero field is lost at $\xi < \xi_\mathrm{c}$, with $\xi_\mathrm{c} = 0.0116$ for $\Gamma_1/|K_1| = 0.5$, and gives way to a highly degenerate classical ground state that can be understood as the classical remnant of the quantum spin liquid found for $\xi = 0$ in numerical studies (``$K\Gamma$SL'')\cite{catuneanu2018, gohlke2018}. The set of classical ground states includes several ferrimagnetic states, including those of the two X domains discussed above. The projection of the total moment onto the $\phi = 90^\circ$ in-plane axis is larger for some of these other ferrimagnetic states, and they are therefore favored for small magnetic fields in this direction. Above some $\xi$-dependent critical field strength, however, we again observe a transition into the intermediate X phase for all $\xi \geq 0$. The X phase therefore adiabatically connects the effective spin model for \rucl\ to the pure Kitaev-$\Gamma$ limit with its topologically-ordered gapped spin-liquid ground state\cite{catuneanu2018, gohlke2018}.


\begin{figure*}[p]
\includegraphics[width=0.68\textwidth]{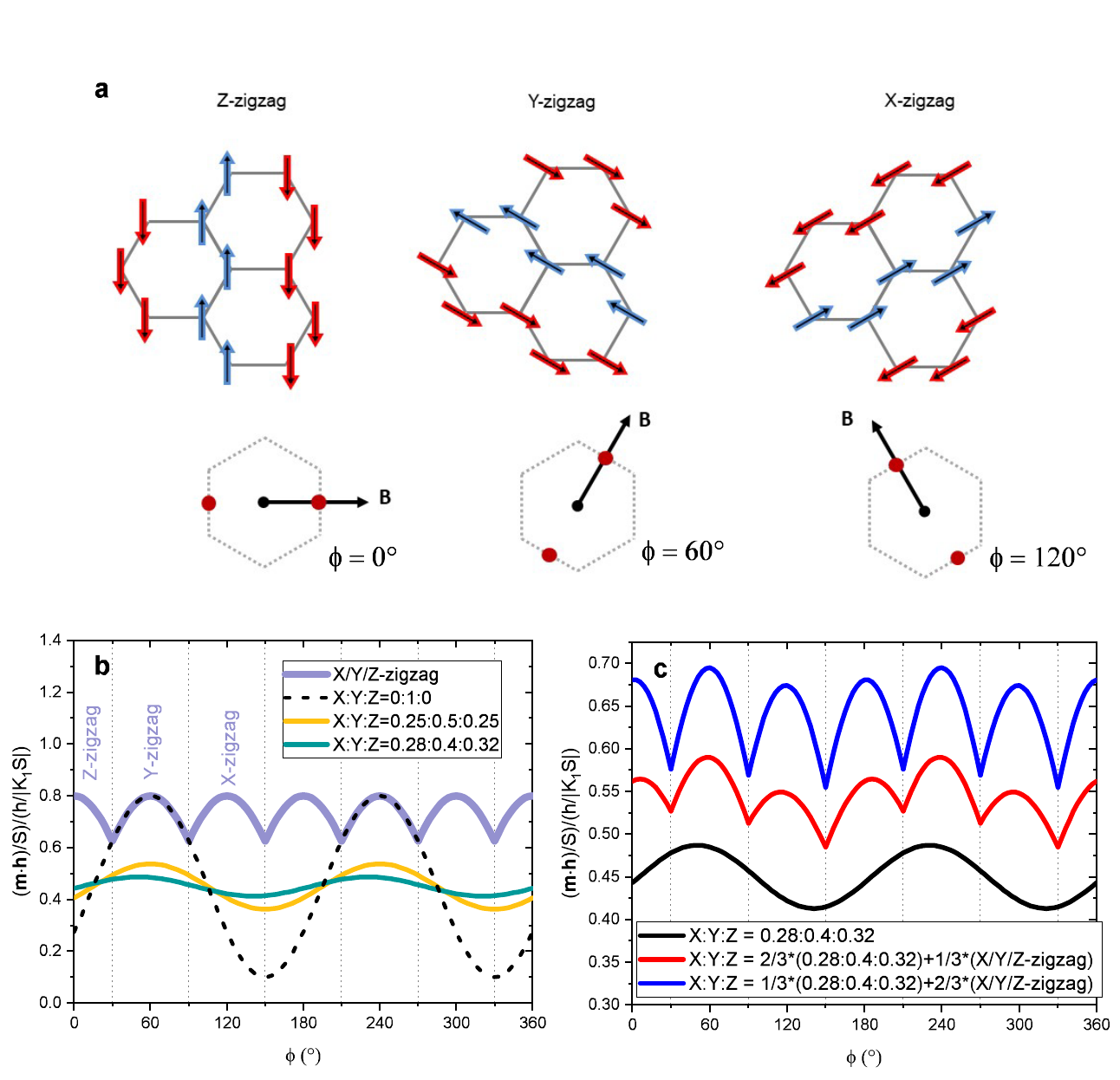}
\caption{
\textbf{Zigzag magnetic domains.} 
(a) Illustration of three possible zigzag magnetic domains in \rucl\ projected in the honeycomb plane (upper panels). The associated magnetic Bragg peak positions (red circles) and magnetic field orientation that stabilizes each domain are illustrated in the reciprocal lattice (grey dashed line) (lower panels).
(b) Expected oscillation of the magnetic susceptibility as a function of magnetic field angle $\phi$ for field-selected domains (purple line) with switching between Z-, Y-, and X-zigzag domains at $\phi=30^{\circ},90^{\circ}, 150^{\circ}$ etc., a single Y-zigzag domain (black dashed line) and unequally populated domains with the fixed ratios X:Y:Z = 0.25:0.5:0.25 (yellow line) and X:Y:Z = 0.28:0.4:0.32 (green line). 
(c) Expected oscillation of the susceptibility as a result of domain repopulation for $B \ll B_\text{dr}$ [black line, same as green line in (b)], $B \lesssim B_\text{dr}$ (red line), and $B \sim B_\text{dr}$ (blue line), where $B_\text{dr}$ is the domain repopulation field.
}
\label{fig:Domains}
\end{figure*}
\begin{figure*}[p]
\includegraphics[width=0.9\textwidth]{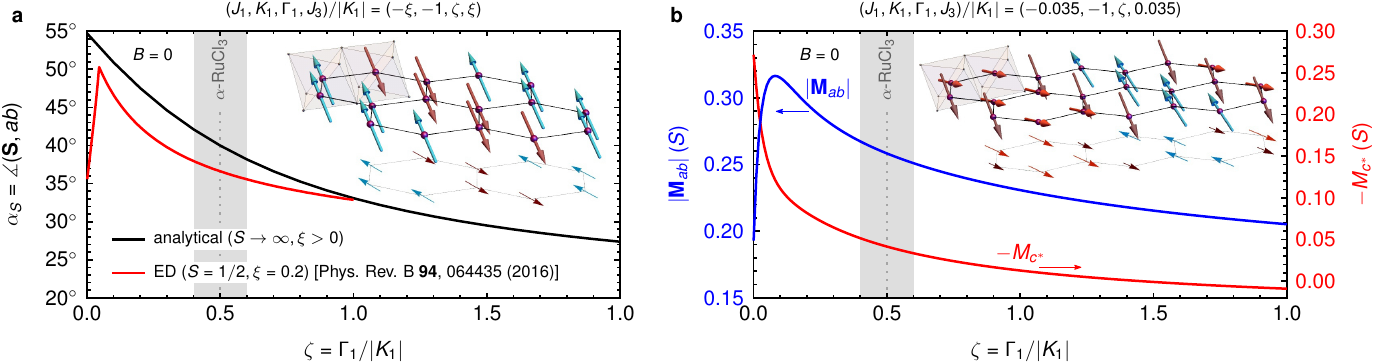}
\caption{
\textbf{Spin orientations} at ${B=0}$.
(a): Angle $\alpha$ between staggered magnetization of ZZ state and honeycomb $ab$ plane as a function of $\zeta = \Gamma_1 / |K_1|$ for $B=0$. The classical result (black) is independent of $\xi = - J_1/|K_1| = J_3 / |K_1|$ and decreases with increasing $\zeta$. Quantum corrections (red, from ref.~\onlinecite{chaloupka2016}) slightly decrease $\alpha$ for the regime relevant for \rucl\ (gray shaded) and introduce a small $\xi$ dependence (here shown: $\xi = 0.2$). Inset: Real-space spin configuration of ZZ state for $\zeta = 0.5$ in the classical limit.
(b): Total magnetization of the (metastable) X state for $B=0$ and $\xi = 0.035$, projected onto the honeycomb $ab$ plane (blue) and the out-of-plane $c^\ast$ axis (red), respectively, as a function of $\zeta$. Inset: Real-space spin configuration of X state for $\zeta = 0.5$.
}
\label{fig:alphaXzeta}
\end{figure*}

\section{Domain selection and oscillations in the zigzag phase}
\label{sec:zz-domains}

We have observed experimentally that the susceptibility of \rucl\ in small magnetic fields $\lesssim 2\ \mathrm{T}$ exhibits a two-fold rather than six-fold oscillation (see Fig.~\ref{fig:SampleCompare}e,f and ref.~\onlinecite{plk2018}). 
This behavior can be understood to originate from zigzag domain repopulation.
The zigzag phase at zero field comprises three domains (we refer to them as X-, Y-, and Z-zigzag), in which the spin direction differs by 120$^{\circ}$. For the model Hamiltonian under consideration, the spins in these domains lie perpendicular to a Ru-Ru bond, i.e., their projections onto the honeycomb plane are along the reciprocal $(1,1,0)$, $(-2,1,0)$, and $(-1,2,0)$ directions, respectively, see Fig.~\ref{fig:Domains}a.
A small external field will favor the domain(s) for which the spin direction is most nearly perpendicular to $\mathbf B$, leading to the largest susceptibility among the different domains.
For a generic angle $\phi$, this selects a unique domain to the detriment of the other two domains, which then become metastable. At the special angles of $\phi \equiv 30^\circ \mod 60^\circ$, however, two domains will be stabilized with respect to the (then metastable) third, which leads to the above-discussed behavior of the magnetic Bragg peaks in an external field.
Within an ideal trigonal environment, and assuming thermal equilibrium at all times, this domain selection would occur already at infinitesimally small fields. 
Small distortions present in \rucl, however, lead to slightly different magnitudes of the Kitaev, Heisenberg, and off-diagonal $\Gamma$ interactions on the three different types of bonds. These bond anisotropies break the $C_3^*$ symmetry of simultaneous threefold rotations in both lattice and spin space of the ideal Heisenberg-Kitaev-$\Gamma$ model.
They will typically favor a particular domain over the other two, and there will generically be a competition between the anisotropy-selected and field-selected zigzag domains at small fields. The domain-repopulation field $B_\text{dr}$, above which the domain is selected by the field direction, will then be shifted to a finite field strength.
In \rucl, this field strength is reported to be of the order of $B_\text{dr} \sim 2$~T (refs.~\onlinecite{sears2017, wolter2017, banerjee2018}). Assuming domain selection by bond anisotropy for fields below $B_\text{dr}$ allows us to estimate the effective energy scale for these anisotropies: $|\Delta J+\Delta K + \Delta \Gamma| \approx g_{ab} \mu_\mathrm{B} B_\text{dr} / S \approx 0.5\ \mathrm{meV}$, which is slightly smaller, but roughly consistent with more elaborate estimates\cite{winter2016, plk2018}.

If a unique domain were selected for $B \ll B_\text{dr}$ throughout the sample, the susceptibility within this single domain would be maximized when the magnetic field is perpendicular to the ordered moment direction in the zigzag chains, generating a two-fold oscillation in the susceptibility as the magnetic field is rotated within the honeycomb plane. This is shown as a dashed black line in Fig.~\ref{fig:Domains}b. 
In general, however, bond anisotropy may select different domains in different spatial regions of the sample, and the resulting susceptibility is a weighted average of the three individual domain responses, resulting in a two-fold oscillation with smaller amplitude and a smaller mean value. The green and yellow lines in Fig.~\ref{fig:Domains}b show the domain-averaged susceptibility for two different unequal (fixed) domain populations.
For $B \gg B_\text{dr}$, and assuming full equilibration, rotation of the magnetic field produces, by contrast, a six-fold oscillation as a result of the periodic switching of the field-selected domain from Z- to Y- to X-zigzag at $\phi=30^{\circ}, 90^{\circ}, 150^{\circ}$, etc.\ (purple line in Fig.~\ref{fig:Domains}b).
For intermediate fields $B \lesssim B_\text{dr}$, we can emulate the domain-selection process phenomenologically by superimposing the two-fold curve originating from anisotropy-selected domains with a six-fold curve arising from field-selected domains. This is shown for two different superpositions (corresponding to two different field strengths) as red and blue curves in Fig.~\ref{fig:Domains}c, with the black curve showing for comparison the two-fold response without a six-fold superposition (same as green curve in Fig.~\ref{fig:Domains}b).
The theoretical curves in Fig.~\ref{fig:Domains}c should be compared with the measured susceptibilities shown in Fig.~\ref{fig:SampleCompare}e for 0.1 T (black curve), 0.5 T (red curve), and 1.0 T (blue curve). We find qualitative agreement.

We note that the measured susceptibilities at higher fields $B > B_\text{dr}$ exhibit a residual two-fold oscillation in addition to the prevailing six-fold oscillation. This effect appears to be somewhat sample dependent (cf.\ Fig.~\ref{fig:SampleCompare}e,f), and we attribute it to bond anisotropies, which change the energetics among the different field-selected domains. The effect of bond anisotropies, including the consequences for the susceptibilities at higher temperatures (Fig.~\ref{fig:SampleCompare}g,h), is discussed in detail in ref.~\onlinecite{plk2018}.


\section{Ordered moments in the $J_1$-$K_1$-$\Gamma_1$-$J_3$ model at zero field}

In order to understand the response of the system to an external magnetic field, it is instructive to discuss the directions of the ordered moments at zero field.
For $\mathbf B = 0$, the Heisenberg-Kitaev-$\Gamma$ model has, for sufficiently large $J_3$, an antiferromagnetic zigzag ground state. For $\Gamma_1/|K_1| \gtrsim 0.05$ (ref.~\onlinecite{chaloupka2016}), the pseudospins are aligned along
\begin{equation}
	\mathbf{S}_i/S = \pm \left[ \mathbf{e}_{\{1,1,0\}} \cos \alpha_S + \mathbf{e}_{c^*} \sin \alpha_S \right],
\end{equation}
where $\alpha_S$ denotes the angle between the pseudospin direction $\mathbf S_i$ and the $ab$ plane, and $\mathbf{e}_{\{1,1,0\}}$ and $\mathbf{e}_{c^*}$ are unit vectors with $\mathbf{e}_{\{1,1,0\}} \parallel \{1,1,0\}$ and $\mathbf{e}_{c^*} \parallel (0,0,1)$.
Fig.~\ref{fig:alphaXzeta}a shows $\alpha_S$ as a function of $\zeta = \Gamma_1/|K_1|$ in the classical limit, $S \to \infty$. In this limit, the pseudospin orientation is independent of the Heisenberg interactions $J_1$ and $J_3$ (ref.~\onlinecite{janssen2017}).
For comparison, we have also included a previous result in the quantum limit for $S=1/2$ that was obtained by exact diagonalization (ED) on a 24-site cluster~\cite{chaloupka2016}. Quantum corrections introduce a weak dependence on the Heisenberg interactions and slightly decrease $\alpha_S$ for the regime relevant for \rucl\ ($\zeta \simeq 0.5$).
From this, we estimate $\alpha_S \simeq + 37^\circ$ for \rucl.
The direction of the magnetic moments as measured in, e.g., neutron experiments, however, differs from the pseudospin direction if the $g$ tensor is anisotropic~\cite{chaloupka2016}. Using the estimate employed in ref.~\onlinecite{winter2018},
\begin{align}
	g =
	\begin{pmatrix}
	g_{ab} & 0 & 0\\
	0 & g_{ab} & 0\\
	0 & 0 & g_{c^*}
	\end{pmatrix}
\simeq
	\begin{pmatrix}
	2.3 & 0 & 0\\
	0 & 2.3 & 0\\
	0 & 0 & 1.3
	\end{pmatrix},
\end{align}
we find for the angle $\alpha_m$ between the magnetic moments $\mathbf m_i = g\,\mathbf S_i$ and the $ab$ plane
\begin{equation}
	\alpha_m = \arccos\left( \frac{g_{ab} \cos \alpha_S}{\sqrt{g_{ab}^2 \cos^2 \alpha_S + g_{c^*}^2 \sin^2 \alpha_S}} \right) \simeq + 23^\circ,
\end{equation}
which is consistent with the latest experimental result for \rucl\ using polarized neutrons~\cite{balz2018}.

Fig.~\ref{fig:alphaXzeta}b shows the total magnetization $\mathbf{M}_\mathrm{tot} = \mathbf{M}_{ab} + M_{c^*} \mathbf{e}_{c^*}$ with $\mathbf{M}_{ab} \perp \mathbf{e}_{c^*}$ in the (then metastable) X state at zero external field as a function of $\zeta = \Gamma_1 / |K_1|$. Its size is indicative of the ferrimagnetic nature of this state. The projection onto the $ab$ plane ($|\mathbf{M}_{ab}|$, blue) for the regime relevant for \rucl\ (gray shaded) is large, while the out-of-plane component ($M_{c^*} $, red) is relatively small in this regime. The X state is therefore favored for in-plane field directions and suitable $\xi$ and $\zeta$, while other states may be stabilized for out-of-plane fields.



\begin{thebibliography}{99}

\bibitem[$\ast$]{contrib} 
P.L.K.\ and L.J.\ contributed equally to this work.

\bibitem[$\dagger$]{plk} 
\href{mailto:kelleypj@ornl.gov}{kelleypj@ornl.gov}

\bibitem[$\ddagger$]{lj} 
\href{mailto:lukas.janssen@tu-dresden.de}{lukas.janssen@tu-dresden.de}

\setcounter{NAT@ctr}{0}

\bibitem{Kit06}
A.~Kitaev, Ann. Phys. (N.Y.) {\bf 321}, 2 (2006).

\bibitem{Jac09}
G.~Jackeli and G.~Khaliullin,
Phys. Rev. Lett. {\bf 102}, 017205 (2009).

\bibitem{Cha10}
J.~Chaloupka, G.~Jackeli, and G.~Khaliullin,
Phys. Rev. Lett. {\bf 105}, 027204 (2010).

\bibitem{plumb14}
K. W. Plumb, J. P. Clancy, L. J. Sandilands, V. V. Shankar, Y. F. Hu, K. S. Burch, H. Y. Kee, and Y. J. Kim,
Phys. Rev. B {\bf 90}, 041112(R) (2014).

\bibitem{sears15}
J. A. Sears, M. Songvilay, K. W. Plumb, J. P. Clancy, Y. Qiu, Y. Zhao, D. Parshall, and Y.-J. Kim,
Phys. Rev. B {\bf 91}, 144420 (2015).

\bibitem{banerjee16}
A. Banerjee, C. A. Bridges, J.-Q. Yan, A. A. Aczel, L. Li, M. B. Stone, G. E. Granroth, M. D. Lumsden, Y. Yiu, J. Knolle, S. Bhattacharjee, D. L. Kovrizhin, R. Moessner, D. A. Tennant, D. G. Mandrus, and S. E. Nagler,
Nat. Mater. {\bf 15}, 733 (2016).

\bibitem{banerjee17}
A. Banerjee, J.-Q. Yan, J. Knolle, C. A. Bridges, M. B. Stone, M. D. Lumsden, D. G. Mandrus, D. A. Tennant, R. Moessner, and S. E. Nagler,
Science {\bf 356}, 1055 (2017).

\bibitem{gohlke17}
M. Gohlke, R. Verresen, R. Moessner, and F. Pollmann,
Phys. Rev. Lett. \textbf{119}, 157203 (2017).

\bibitem{wolter17}
A. U. B. Wolter, L. T. Corredor, L. Janssen, K. Nenkov, S. Sch\"onecker, S.-H. Do, K.-Y. Choi, R. Albrecht, J. Hunger, T. Doert, M. Vojta, and B. B\"uchner,
Phys. Rev. B \textbf{96}, 041405(R) (2017).

\bibitem{leahy17}
I. A. Leahy, C. A. Pocs, P. E. Siegfried, D. Graf, S.-H. Do, K.-Y. Choi, B. Normand, and M. Lee,
Phys. Rev. Lett. {\bf 118}, 187203 (2017).

\bibitem{baek17}
S.-H. Baek, S.-H. Do, K.-Y. Choi, Y. S. Kwon, A. U. B. Wolter, S. Nishimoto, J. van den Brink, and B. B\"uchner,
Phys. Rev. Lett. \textbf{119}, 037201 (2017).

\bibitem{sears17}
J. A. Sears, Y. Zhao, Z. Xu, J. W. Lynn, and Young-June Kim,
Phys. Rev. B {\bf 95}, 180411(R) (2017).

\bibitem{zheng17}
J. Zheng, K. Ran, T. Li, J. Wang, P. Wang, B. Liu, Z. Liu, B. Normand, J. Wen, and W. Yu,
Phys. Rev. Lett. \textbf{119}, 227208 (2017).

\bibitem{winter18}
S. M. Winter, K. Riedl, D. Kaib, R. Coldea, and R. Valent{\'i},
Phys. Rev. Lett. \textbf{120}, 077203 (2018).

\bibitem{hentrich18}
R. Hentrich, A. U. B. Wolter, X. Zotos, W. Brenig, D. Nowak, A. Isaeva, T. Doert, A. Banerjee, P. Lampen-Kelley, D. G. Mandrus, S. E. Nagler, J. Sears, Y.-J. Kim, B. B\"uchner, and C. Hess,
Phys. Rev. Lett. \textbf{120}, 117204 (2018).

\bibitem{banerjee2018}
A. Banerjee, P. Lampen-Kelley, J. Knolle, C. Balz, A. A. Aczel, B. Winn, Y. Liu, D. Pajerowski, J.-Q. Yan, C. A. Bridges, A. T. Savici, B. C. Chakoumakos, M. D. Lumsden, D. A. Tennant, R. Moessner, D. G. Mandrus, and S. E. Nagler,
npj Quantum Mater. {\bf 3}, 8 (2018).

\bibitem{Matsuda18}
Y. Kasahara, T. Ohnishi, Y. Mizukami, O. Tanaka, S. Ma, K. Sugii, N. Kurita, H. Tanaka, J. Nasu, Y. Motome, T. Shibauchi, and Y. Matsuda,
arXiv:1805.05022 (unpublished).

\bibitem{Kimchi11}
I.~Kimchi and Y.-Z.~You,
Phys. Rev. B {\bf 84}, 180407 (2011).

\bibitem{Cha13}
J.~Chaloupka, G.~Jackeli, and G.~Khaliullin,
Phys. Rev. Lett. {\bf 110}, 097204 (2013).

\bibitem{rau14}
J. G.~Rau, E. K.-H.~Lee, and H-Y.~Kee,
Phys. Rev. Lett. {\bf 112}, 077204 (2014).

\bibitem{perkins14}
Y. Sizyuk, C. Price, P. W\"olfle, and N. B. Perkins,
Phys. Rev. B {\bf 90}, 155126 (2014).

\bibitem{ioannis15}
I. Rousochatzakis, J. Reuther, R. Thomale, S. Rachel, and N. B. Perkins,
Phys. Rev. X {\bf 5}, 041035 (2015).

\bibitem{winter16}
S. M. Winter, Y. Li, H. O. Jeschke, and R. Valent{\'i},
Phys. Rev. B {\bf 93}, 214431 (2016).

\bibitem{winter17}
S. M. Winter, K. Riedl, A. Honecker, and R. Valent\'i,
Nat. Commun. \textbf{8}, 1152 (2017).

\bibitem{janssen17}
L. Janssen, E. C. Andrade, and M. Vojta,
Phys. Rev. B \textbf{96}, 064430 (2017).

\bibitem{Yu18}
Y. J. Yu, Y. Xu, K. J. Ran, J. M. Ni, Y. Y. Huang, J. H. Wang, J. S. Wen, and S. Y. Li, 
Phys. Rev. Lett. {\bf 120}, 067202 (2018).

\bibitem{catun18}
A. Catuneanu, Y. Yamaji, G. Wachtel, Y. B. Kim, and H.-Y. Kee,
npj Quantum Mater.\ {\bf 3}, 23 (2018).

\bibitem{gohlke18}
M. Gohlke, G. Wachtel, Y. Yamaji, F. Pollmann, and Y. B. Kim,
Phys. Rev. B \textbf{97}, 075126 (2018).

\bibitem{rousochatzakis2017}
I. Rousochatzakis and N. B. Perkins,
Phys. Rev. Lett. {\bf 118}, 147204 (2017).

\bibitem{agrestini2017}
S. Agrestini, C.-Y. Kuo, K.-T. Ko, Z. Hu, D. Kasinathan, H. B. Vasili, J. Herrero-Martin, S. M. Valvidares, E. Pellegrin, L.-Y. Jang, A. Henschel, M. Schmidt, A. Tanaka, and L. H. Tjeng,
Phys. Rev. B {\bf 96}, 161107(R) (2017).

\bibitem{plk2018}
P. Lampen-Kelley, S. Rachel, J. Reuther, J.-Q. Yan, A. Banerjee, C. A. Bridges, H. B. Cao, S. E. Nagler, and D. Mandrus,
arXiv:1803.04871.

\end{thebibliography}

\begin{thebibliography}{99}

\bibitem[$\ast$]{contrib} 
P.L.K.\ and L.J.\ contributed equally to this work.

\bibitem[$\dagger$]{plk} 
\href{mailto:kelleypj@ornl.gov}{kelleypj@ornl.gov}

\bibitem[$\ddagger$]{lj} 
\href{mailto:lukas.janssen@tu-dresden.de}{lukas.janssen@tu-dresden.de}

\setcounter{NAT@ctr}{0}

\bibitem{Johnson2015}
R. D. Johnson, S. C. Williams, A. A. Haghighirad, J. Singleton, V. Zapf, P. Manuel, I. I. Mazin, Y. Li, H. O. Jeschke, R. Valent\'{i}, and R. Coldea,
Phys. Rev. B {\bf 92}, 235119 (2015).

\bibitem{Cao2016}
H. B. Cao, A. Banerjee, J.-Q. Yan, C. A. Bridges, M. D. Lumsden, D. G. Mandrus, D. A. Tennant, B. C. Chakoumakos, and S. E. Nagler,
Phys. Rev. B {\bf 93}, 134423 (2016).

\bibitem{plk2018}
P. Lampen-Kelley, S. Rachel, J. Reuther, J.-Q. Yan, A. Banerjee, C. A. Bridges, H. B. Cao, S. E. Nagler, and D. Mandrus,
arXiv:1803.04871.

\bibitem{banerjee2018}
A. Banerjee, P. Lampen-Kelley, J. Knolle, C. Balz, A. A. Aczel, B. Winn, Y. Liu, D. Pajerowski, J.-Q. Yan, C. A. Bridges, A. T. Savici, B. C. Chakoumakos, M. D. Lumsden, D. A. Tennant, R. Moessner, D. G. Mandrus, and S. E. Nagler, 
%
npj Quantum Mater. {\bf 3}, 8 (2018).

\bibitem{catuneanu2018}
A. Catuneanu, Y. Yamaji, G. Wachtel, Y. B. Kim, and H.-Y. Kee,
npj Quantum Mater. {\bf 3}, 23 (2018).

\bibitem{gohlke2018}
M. Gohlke, G. Wachtel, Y. Yamaji, F. Pollmann, and Y. B. Kim,
Phys. Rev. B \textbf{97}, 075126 (2018).

\bibitem{chaloupka2016}
J. Chaloupka and G. Khaliullin,
Phys. Rev. B {\bf 94}, 064435 (2016).

\bibitem{sears2017}
J. A. Sears, Y. Zhao, Z. Xu, J. W. Lynn, and Young-June Kim, 
Phys. Rev. B {\bf 95}, 180411(R) (2017).

\bibitem{wolter2017}
A. U. B. Wolter, L. T. Corredor, L. Janssen, K. Nenkov, S. Sch\"onecker, S.-H. Do, K.-Y. Choi, R. Albrecht, J. Hunger, T. Doert, M. Vojta, and B. B\"uchner, 
Phys. Rev. B {\bf 96}, 041405(R) (2017).

\bibitem{winter2016}
S. M. Winter, Y. Li, H. O. Jeschke, and R. Valent\'i, 
Phys. Rev. B {\bf 93}, 214431 (2016).

\bibitem{janssen2017}
L. Janssen, E. C. Andrade, and M. Vojta,
Phys. Rev. B {\bf 96}, 064430 (2017).

\bibitem{winter2018}
S. M. Winter, K. Riedl, D. Kaib, R. Coldea, and R. Valent\'i,
Phys. Rev. Lett. {\bf 120}, 077203 (2018).

\bibitem{balz2018}
C. Balz {\it et al.}, unpublished.

\end{thebibliography}
\end{document}